%% file: fide-user-assistance.tex
\documentclass[copyright]{eptcs}
\providecommand{\event}{F-IDE 2016} 
\usepackage{underscore}             

\usepackage[utf8]{inputenc}
\usepackage{fixltx2e}
\usepackage[T1]{fontenc}
\usepackage[english]{babel}
\usepackage{graphicx}
\usepackage{url}
\input{extra/listings}

\title{User Assistance Characteristics \\ of the USE Model Checking Tool}
\author{Frank~Hilken \qquad\qquad Martin~Gogolla
\institute{University of Bremen, Computer Science Department \\ 28359 Bremen, Germany}
\email{\{fhilken,gogolla\}@informatik.uni-bremen.de}}
\def\titlerunning{User Assistance Characteristics of the USE Model Checking Tool}
\def\authorrunning{F.~Hilken \& M.~Gogolla}

\begin{document}
\maketitle

\begin{abstract}
The Unified Modeling Language~(UML) is a widely used general purpose
modeling language. Together with the Object Constraint Language~(OCL),
formal models can be described by defining the structure and behavior
with UML and additional OCL constraints. In the development process
for formal models, it is important to make sure that these models are
(a)~correct, i.e.~consistent and complete, and (b)~testable in the
sense that the developer is able to interactively check model
properties. The USE tool~(UML-based Specification Environment) allows
both characteristics to be studied. We demonstrate how the tool
supports modelers to analyze, validate and verify UML and OCL models
via the use of several graphical means that assist the modeler in
interpreting and visualizing formal model descriptions. In particular, we
discuss how the so-called USE model validator plugin is integrated into the USE
environment in order to allow non domain experts to use it and construct object
models that help to verify properties like model consistency.
\end{abstract}

\section{Introduction}

Model-Driven Engineering~(MDE) is an approach to software development
concentrating on models in contrast to traditional code-centric
development approaches. Within MDE, models are frequently formulated
in the Unified Modeling Language~(UML) with accompanying formal
restrictions expressed in the Object Constraint Language~(OCL). UML
models are visually specified with several diagram kinds emphasizing
structural and behavioral system aspects. Visual model descriptions
offer a great potential for a user-friendly development
process. Naturally, tools must take up the challenge and provide
interfaces that support the developer in an easygoing way.

The present paper studies formal system descriptions employing UML
class diagrams that are restricted by OCL invariants. The feature set
of UML class diagrams that is handled here and the employed OCL
elements pose a formal semantics. We regard this combination of UML
and OCL as a formal method. The aim of this contribution is to
demonstrate how the development of formal UML and OCL models can be
supported by a user-friendly interface. Employing this interface it is
possible to verify properties like model consistency.


\section{Preliminaries}

\subsection{Running Example in UML and OCL}

\begin{figure}[tp]
  \centering
  \includegraphics[width=\linewidth]{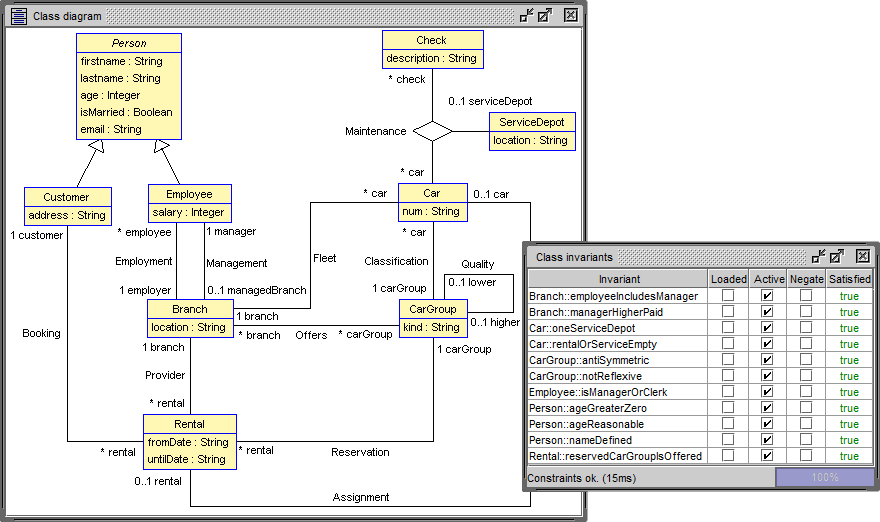}
  \caption{Car rental running example class diagram (left) and invariants (right).}
  \label{fig:runnning-example}
\end{figure}

In UML, class diagrams describe the structure of models with classes and class
attributes, which are templates for 'things' and their properties, e.g. persons
and their personal information. Additionally, associations put the classes in
relation with each other. Figure~\ref{fig:runnning-example} (left) shows the
running example model description. It shows a \emph{Car Rental} model with cars
that are assigned to branches and their maintenance history. Additionally, there
is a categorization for the cars into car groups. Finally, customers can rent
cars from the branches that are run by their employees. The model uses a wide
variety of UML features.

The class diagram is instantiated to create actual scenarios to describe car
rentals. These system states are represented by UML object diagrams. They are
restricted by the semantics of the class diagram and must satisfy all model
inherent constraints given by, e.g.~generalizations, multiplicities or
compositions (not present in this model).

In addition to the UML descriptions, OCL invariants are used to employ further
restrictions on the model that are not expressible by UML alone. These
constraints are pictured in Fig.~\ref{fig:runnning-example} on the right. They
handle further relations between classes and ensure that the model does not
allow system states that are not intended. For example, in the car rental model,
the categorization of the car groups shall be cycle free and all employees must
be connected to a branch, which cannot be handled by multiplicities, because
there are multiple ways to represent this relation (\texttt{Employment} and
\texttt{Management}).

The goal is to create a model description that can represent all intended
situations but not more. Using model checking tools, the models can then be
checked for certain properties. Usually, these properties regard safety, but
other concerns can be checked as well. A valid system state must satisfy all
model inherent constraints as well as all concrete constraints given by the OCL
invariants.

\subsection{Model Verification with the USE Tool}

In this paper, the \emph{UML-based Specification Environment}
tool~(USE,~\cite{Gogolla:2007:SCP}) is used together with the so-called
\emph{model validator} plugin that allows to generate system states for UML/OCL
models based on a relational logic encoding~\cite{Kuhlmann:2012:MODELS}. There is
earlier work showing features of the USE tool that help to analyze and debug OCL
expressions, namely the evaluation browser~\cite{Bruening:2012:TAP}, but we
concentrate on the model validator and its model checking aspects in this work.

In order for the model validator to search for a valid system state, it needs
several inputs:
\begin{itemize}
  \item A description of the \emph{model} in the form of a class diagram optionally
  enhanced with invariants given as OCL expressions, see
  Fig.~\ref{fig:runnning-example}.
  \item A \emph{configuration}, which defines the search space by providing the domain
  of basic data types (Integer, String and Real) and lower and upper bounds for
  classes and associations, i.e. specifies how many objects of each class -- or
  links of each association respectively -- are required and how many are
  allowed at most. Furthermore, the configuration contains rules for the
  assignments of class attributes, e.g. restricting domain values for certain
  attributes specifically. Finally, it allows to disable or negate invariants.
  \item Optionally a \emph{partial system state} can be instantiated before the
  validation process that is used as a base for the task. The model validator
  adds elements to the system state until it: (a)~conforms the bounds given in
  the configuration; and (b)~is a valid system state as defined by the model.
  This can be seen as a lower bound on the model level.
  \item Smaller, model independent parameters include the choice of the
  \emph{SAT solver} and \emph{bitwidth} for the encoding of the model.
\end{itemize}

The second bullet point, the configuration, previously required the modeler to
edit a text file containing key-value pairs to setup values for certain keys.
The keys are determined by the model. For example for each class and association
a lower and an upper bound is expected. The required values are mostly numbers,
but more complex constructs were required to specify, e.g.~preexisting links.
This process requires a deep understanding of the existing syntax
to enter the values. Moreover, special values exist for some keys with different
meanings, e.g. unlimited upper bounds. Even experienced users regularly required
the manual of the configuration.

In the following sections, we explain how a new graphical user interface helps
to simplify the configuration process and reduce the necessity for a separate
manual to the tool. Furthermore, additional analysis features of the tool are
presented to help identify potential problems with the verification process
caused by the inputs.

\section{Iterative Instantiation of the Running Example}
\label{sec:conf-construction}

We now study for our running example four use cases corresponding to four model
validator configurations that result in UML object diagrams. The basic structure
of the GUI contains three tabs for (1)~the datatypes, (2)~the classes and
associations, and (3)~the invariants. We iteratively build up the configuration
to generate multiple object diagrams.

\begin{description}
\item[Datatypes] First, only basic OCL types (e.g.~\texttt{Integer}
and~\texttt{String}) are configured by giving bounds for their domain. For
example, it is determined that integers may range from $-10$ to $10$ and we want
to use at most $10$ string values, which are automatically generated.

\item[Parts of the Model] In the next step, the bounds for some of the classes of
the model are configured and their relevant invariants are enabled.
Figure~\ref{fig:1_caseB_SC} shows the configuration tabs and the resulting
object diagram for this step. Here, the bounds for the
classes~\texttt{Customer}, \texttt{Employee} and~\texttt{Branch} are set to
\texttt{1..1} and all others are set to \texttt{0}. The class \texttt{Person} is
abstract and, therefore, cannot be setup. All invariants based on these classes
are enabled as well and none are negated. These settings are useful to setup
certain model verification tasks, e.g.~invariant
independence~\cite{DBLP:conf/tap/GogollaKH09}. In addition, the
associations~\texttt{Management} and~\texttt{Employment} are enabled with a
bound configuration of \texttt{1..1}. The object diagram shows the default
string values for attributes, since these have not been specified further.

\begin{figure}[p]
  \centering
  \includegraphics[width=.945\linewidth]{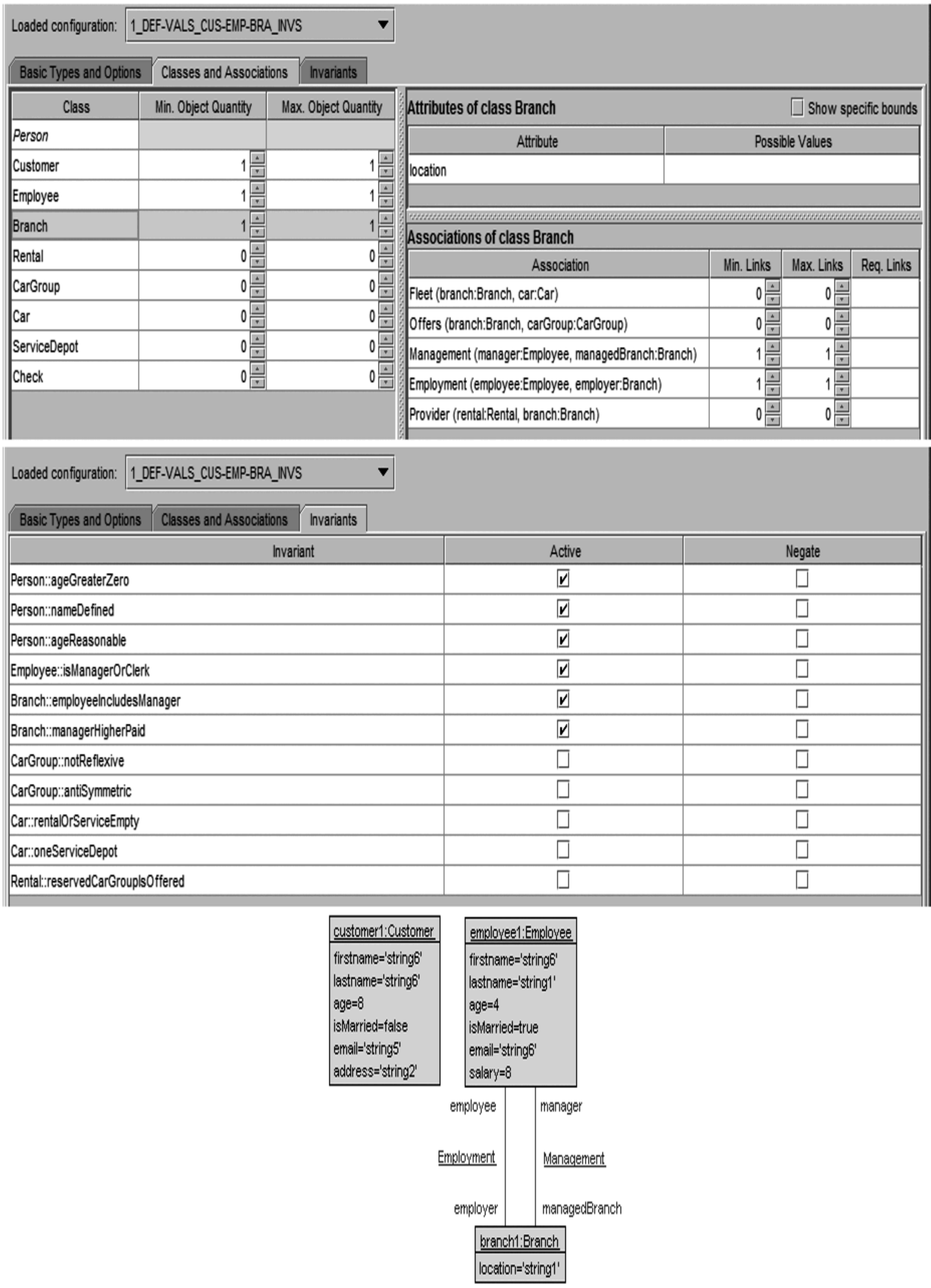}
  \caption{Configuration of parts of the model considering only a few classes and associations.}
  \label{fig:1_caseB_SC}
\end{figure}

\item[Full Model] Once parts of the model have successfully been instantiated, a
configuration enabling all elements is built and run through the model
validator. With these smaller steps per configuration, there is less margin for
errors and if there is one, it is easier detectable.

\item[Application Specific Values] Finally, application specific datatype values
are employed for class attributes and basic types. They lead in the constructed
object diagram to a state that seems more realistic, more domain specific than
the previous object diagram. This also allows to specify that , e.g.~address
strings are not used for names.
\end{description}

\section{Configuration GUI}

First off, with the release of the configuration GUI the interface of
configuration files was extended to allow storing multiple configurations in one
file. These configurations can be named individually and the configuration GUI
offers operations to manage them, e.g. cloning configurations, renaming or deleting them etc. This allows
for an easy iterative construction of the configurations as shown in
Sect.~\ref{sec:conf-construction}. The GUI also offers common file system
operations to deal with the generated configuration files. For convenience, a
configuration file with the same base name as the loaded model is automatically
opened if one exists.

The GUI is split into three tabs that each cover a part of the
configuration.\footnote{A detailed explanation of all three tabs in detail
including screenshots can be found in~\cite{fide16-longversion}.} The first tab
is the basic types tab in which the domains for the basic types of OCL are
defined. The domains can be defined as ranges or specific values that will be
used by the model validator.

The second tab defines the model dependent bounds and domains. These include
bounds for classes, attributes and associations as seen in the top of
Fig.~\ref{fig:1_caseB_SC}. The GUI only requires the values for the specific
settings which eliminates the need to know the syntax for each model element,
which makes the creation of configurations simpler and faster. Further, abstract
classes cannot be setup, which is represented with non-editable fields in the
GUI. If a value cannot be parsed, it is highlighted red and the modeler
immedietly sees problems in the configuration. Features that require expert
knowledge are hidden behind a checkbox.

The final, third tab, configures the invariants. Invariants can be individually
deactivated and negated, which is required for certain verification
tasks~\cite{DBLP:conf/tap/GogollaKH09}.

\section{Analysis of Potential Modeling Problems}

So far, we have shown how the user is assisted by the graphical user interface
to setup the configuration of a verification task. However, besides a bad
configuration, other problems can interfere with the checking process, in
particular problems that the user is not aware of.

UML and OCL are rich languages filled with features for all kinds of purposes.
Trying to support all of them is not only a lot of work, but also reduces the
efficiency of the tools, the more features they
support~\cite{DBLP:conf/tap/HilkenNGW14}. Therefore, it is common practice to
restrict UML and OCL verification engines to a subset of the languages. This
results in some features being completely unsupported and others only having
limited support. Both categories pose problems to the users of the tools. If
there are no means in place to detect the limitations, the outcome might differ
from the user's expectations and it is not feasible to keep track of all
limitations from long tool manuals.

The verification engine of the USE tool, the model validator, has good support
for UML and OCL, but also has limits. The
underlying solving engine of the model validator is based on relational logic,
which has great support for set operations and, thus, the integration of the OCL
\texttt{Set} collection type is extensive. Adding support for the other
collection types~\texttt{Bag}, \texttt{Sequence} and \texttt{OrderedSet} would
pose a significant overhead though, i.e.~will be less efficient. This
restriction to the \texttt{Set} collection type is particularly problematic for
OCL navigation expressions. This expression allows to navigate the classes of
the model using associations, more precisely the roles visible in
Fig.~\ref{fig:runnning-example}. Usually a 1-$n$ navigation results in a set of
elements, because at most one link is allowed between two objects, but under
certain circumstances -- namely starting with a set of objects rather than a
single one -- the navigation results in the duplicate preserving \texttt{Bag}
type, because the result might contain the same value multiple times after the
navigation. Due to the implicit nature of
this effect and the strict interpretation of bags as sets in the model
validator, simple expressions might already suffer unintended side effects. To
help identify those potential problems, the occurrences are made visible to the
modeler via a warning.

\begin{lstlisting}[style=verbatim]
WARNING: Collect operation `[...].employee.age' results in unsupported type `Bag'. It will be interpreted as `Set'.
\end{lstlisting}

The implicit type change from \texttt{Set} to \texttt{Bag} in OCL, which is
consequently interpreted as \texttt{Set} by the model validator, brings more
potential problems with it. Most OCL collection operations are defined on all
collection types, but the results are different. Consider the
operation~\texttt{sum()}, which sums all integer elements of a collection. Here, the implicit conversion from \texttt{Set} to
\texttt{Bag} is usually helpful when collecting, for example, the ages of
persons to calculate an average. However, the interpretation as a \texttt{Set}
does not work in this situation. To assist the user, the model validator checks
for these situations and warns the modeler about this potential problem, which
only the modeler can decide whether it needs to be addressed or not.

\begin{lstlisting}[style=verbatim]
WARNING: The evaluation of sum expression `[...].employee.age->sum()' might be wrong if source contains duplicates (Collection is interpreted as Set).
\end{lstlisting}

Other problems might arise from contradictions in the model itself. Navigations
through the model can become quite long and obscure the resulting
type. In these situations, typecasts like~\texttt{oclAsSet()} need to be used to
be able to compare differing types, but the textual representation of OCL alone
is often insufficient to recognize such disparities. USE is able to structurally
analyze the expressions in the model for type contradictions and gives hints
about (sub)expressions that were determined to be contradicting, resulting in
constant values.

\begin{lstlisting}[style=verbatim]
WARNING: Expression `Set{ 1 } = Bag{ 1 }' can never evaluate to true because `Set(Integer)' and `Bag(Integer)' are unrelated.
\end{lstlisting}

Finally, the underlying solving engine is bound to a bitwidth that has to
be specified with the verification task. Setting the bitwidth as small as
necessary increases the efficiency of the tool, thus we leave the task to the
user to choose an appropriate bitwidth. But if the bitwidth is chosen too small,
undefined behavior occurs when dealing with arithmetic operations exceeding the
bitwidth. Finally, if the user is not aware of -- or forgets -- that the
bitwidth is specified in two's-complement, off-by-one errors can occur.
In order to alleviate the problem, the model validator analyzes the
configuration and model description for integer
literals and checks them against the given bitwidth. If it is determined that
the chosen bitwidth is too small, a warning is displayed including the
appropriate bitwidth for the current model and configuration.

\begin{lstlisting}[style=verbatim]
WARNING: The configured bitwidth is too small for the property Integer max value (237). Required bitwidth: 9 or greater.
\end{lstlisting}

\section{Conclusion and Future Work}
\label{sec:conclusion}

We have presented the USE tool together with its model validator plugin and have
shown the steps necessary to apply model checking to given UML/OCL models.
Furthermore, the simplifications of the process by integrating the graphical user
interface have been discussed and the possibilities of the configuration GUI and
the coverage mode are demonstrated. Finally, the possibilities of the model
validator plugin to detect potential problems have been demonstrated to guide
users in finding incompatibilities in their models.

Besides the configuration of the model domains and bounds, we have presented
more aspects that have to be setup before a system state can be generated
including the verification task itself, e.g.~by manipulating the invariants.
Future work should concentrate on the simplification of all steps of the setup
and provide easy interfaces for each of them.
Additionally, the evaluation of the configuration GUI and other presented
interfaces is an ongoing process and new assistance features are constantly
added and improved.

\paragraph{Acknowledgement.}
We thank Subi~Aili for his contributions to the configuration GUI -- ideas and
implementation -- in his diploma thesis.

\bibliographystyle{eptcs}
\bibliography{references,fide2016sub,ag}

\end{document}

%% file: extra/listings.tex
\usepackage{listings}

\lstdefinelanguage{umlocl}{%
morekeywords={%
model,abstract,class,associationclass,association,composition,aggregation,between,role,ordered,subsets,redefines,attributes,constraints,inv,operations,pre,post,end,%
Real,String,Integer,Boolean,enum,%
from,into,%
self,context,package,endpackage,body,init,derive,def,static,let,in,oclAsType,oclIsTypeOf,oclIsKindOf,oclInState,oclIsNew,allInstances,%
OclInvalid,OclVoid,OclMessage,OclAny,UnlimitedNatural,true,false,null,invalid,if,then,else,endif,concat,size,substring,%
Collection,Set,OrderedSet,Sequence,Bag,Tuple,%
oclIsUndefined,oclIsInvalid,oclType,oclLocale,hasReturned,return,isSignalSent,isOperationCall,%
abs,floor,round,max,min,toString,div,mod,size,concat,substring,toInteger,toReal,toUpperCase,toLowerCase,indexOf,equalsIgnoreCase,at,characters,toBoolean,and,or,xor,not,implies,%
select,reject,collect,forAll,exists,closure,iterate,includes,excludes,count,includesAll,excludesAll,isEmpty,notEmpty,sum,product,asSet,asOrderedSet,asSequence,asBag,flatten,union,intersection,including,excluding,symmetricDifference,append,prepend,insertAt,subOrderedSet,first,last,reverse,isUnique,any,one,collectNested,sortedBy,%
isDefined,isUndefined,oclUndefined,%
begin,delete,create,destroy,set%
},%
otherkeywords={!create,!destroy,!insert,!delete,!set,!openter,!opexit,@pre},%
morestring=[b]',%
morecomment=[l]{--},%
morecomment=[l]{//},%
commentstyle=\slshape,%
literate={->}{$\rightarrow$}2%
{ä}{{\"a}}1{ö}{{\"o}}1{ü}{{\"u}}1{Ä}{{\"A}}1{Ö}{{\"O}}1{Ü}{{\"U}}1{ß}{{\ss}}1%
}

\makeatletter
\lstdefinestyle{umlocl}{
language=umlocl,
basicstyle=\upshape\ttfamily\lst@ifdisplaystyle\small\fi,
basewidth={.5em, .5em}
}

\lstdefinestyle{pseudo}{
basicstyle=\upshape\ttfamily\lst@ifdisplaystyle\small\fi,
basewidth={.5em, .5em},
morecomment=[l]{;},
morecomment=[l]{//},
commentstyle=\slshape
}

\lstdefinestyle{numbered}{
numbers=left,
numberstyle=\scriptsize,
xleftmargin=2em,
framexleftmargin=2em
}

\lstdefinestyle{verbatim}{
basicstyle=\upshape\ttfamily\lst@ifdisplaystyle\small\fi,
basewidth={.5em, .5em}
}
\makeatother